\documentclass[prb,aps,floats,floatfix,twocolumn,showpacs,superscriptaddress,reprint]{revtex4-1}
\usepackage[USenglish]{babel}
\usepackage{amsmath,amssymb,amsfonts}
\usepackage{graphicx}
\usepackage{dcolumn}
\usepackage{bm}
\usepackage{epstopdf}
\usepackage[colorlinks=true,linkcolor=blue,citecolor=red]{hyperref}
\usepackage[T1]{fontenc} 
\usepackage{lmodern}
\usepackage[usenames,dvipsnames]{color}
\newcommand{\rozmiarjeden}{0.48\textwidth}
\newcommand{\rozmiardwa}{0.48\textwidth}
\newcommand{\rozmiartrzy}{0.465\textwidth}
\newcommand{\QFCOI}{QCOI}
\newcommand{\HFCOI}{HCOI}
\begin{document}

\title{%
Doping driven metal-insulator transitions and charge orderings\\
in the extended Hubbard model
}%

\author{K.~J.~Kapcia}
\email[corresponding author; e-mail:]{konrad.kapcia@ifpan.edu.pl}
\affiliation{Institute of Physics, Polish Academy of Sciences,
  Aleja Lotnik\'ow 32/46, PL-02668 Warszawa, Poland}

\author{S.~Robaszkiewicz}
\affiliation{Faculty of Physics, Adam Mickiewicz University in
  Pozna\'n, ul. Umultowska 85, PL-61614 Pozna\'n, Poland}

\author{M.~Capone}
\affiliation{Scuola Internazionale Superiore di Studi Avanzati
  (SISSA), Via Bonomea 265, I-34136 Trieste, Italy}

\author{A.~Amaricci}
\affiliation{Scuola Internazionale Superiore di Studi Avanzati  and
  Democritos National Simulation Center, Consiglio Nazionale delle
  Ricerche, Istituto Officina dei Materiali, Via Bonomea 265, I-34136
  Trieste, Italy}

\date{November 10, 2016}

\begin{abstract}
We perform a thorough study of an extended Hubbard model featuring
local and nearest-neighbor Coulomb repulsion. Using dynamical
mean-field theory we investigated the  zero temperature
phase-diagram of this model as a function of the chemical doping.
The interplay between local and non-local interaction drives  a variety of
phase-transitions connecting two distinct charge-ordered insulators,
i.e., half-filled and quarter-filled, a charge-ordered metal and a
Mott insulating phase.
We characterize these transitions and the relative stability of the solutions and
we show that the two interactions conspire to stabilize the
quarter-filled charge ordered phase.
\end{abstract}



\maketitle


\section{Introduction}

Strongly correlated materials are characterized by the relevance of the Coulomb interactions
which competes with the kinetic energy, leading to a tendency towards localization of the
carriers~\cite{Fulde2006,Anisimov2010}.
The simplest theoretical description of this competition is obtained
from the Hubbard model~\cite{Hubbard1963} in terms of conduction band electrons
experiencing a {\it local} screened Coulomb repulsion.
Despite its simplicity, the approximate solutions of this model
revealed an incredibly rich physics which has been the object of extensive investigations (e.g. Refs.~\onlinecite{PennPR1966,SpalekPRL1989,Georges1996RMP,Gebhard1997,Montorsi1992,VucicevicPRL2015}).
This model is also the starting point to take into
account other important effects by including additional interactions,
e.g. phonon coupling, orbital ordering or longer-range interaction.

A great deal of attention has been devoted to understand the
effects of non-local short-range electron-electron repulsion, which favors
a spatial charge ordering~\cite{Solyom.2014}.
The possible existence of inhomogeneous distribution of charges was firstly
predicted in two dimensional (2D) electron gas~\cite{WignerPR1934}, as a result of the tendency
to form a Wigner crystal as soon as the energy gain from the
electronic localization tendency exceeds that in the kinetic energy for
an homogeneous electron distribution. This possibility has
been experimentally realized in various semiconductor
structures~\cite{AndreiPRL1988,HaneinPRL1998,KravchenkoRPP2004}.
The effective importance of the electronic interaction
for the 2D electrons gas, e.g. layers of liquid Helium,
has been recently reconciled with the original Wigner crystallization
scenario~\cite{pankovprb2008,CamjayiNP2008,AmaricciPRB2010}.
However other, and somehow more conventional, examples of materials in
which charge order interplays with Mott physics can be found in
narrow-band correlated systems such as transition-metal
dichalcogenides~\cite{MorosanNP2006,NovelloPRB2015},
or other oxides (e.g. manganites, nickelates, cuprates, bismuthates, and
cobaltates)~\cite{ImadaRMP1998,Renner2002,GotoAP2003,DagottoPhysRep2001,NetoScience2015}
as well as low-dimensional organic conductors~\cite{SeoCR2004,JeromeCR2004,Bourbonnais2008} and
heavy-fermion systems~\cite{OchiaiJPSJ1990,FuldeEPL1995}.

From a theoretical perspective all these evidences motivated a careful
analysis of the Extended Hubbard Model (EHM), i.e. a Hubbard model supplemented
with {\it non-local} density-density interaction term. The direct
competition of local and non-local interactions in the EHM captures
both the effects of strong correlations and the tendency of the system
to form inhomogeneous charge distributions. The EHM has been
extensively studied in many different regimes, e.g. strong coupling
limit, quarter- and half-filling, and by means of different
methods~\cite{RobaszkiewiczAPPA1974,LinCJP2000,hoangjpcm2002},
such as Hartree-Fock mean-field~\cite{RobaszkiewiczPSSB1973,RobaszkiewiczAPPA1974,MicnasRMP1990,RosciszewskiJPCM2003,hoangjpcm2002},
Monte Carlo simulations \cite{HirschPRL1984,LinPRB1986,ClayPRB1999},
variational based cluster perturbation theory~\cite{AichhornPRB2004},
lattice exact diagonalization~\cite{PencPRB1994,CalandraPRB2002,MerinoPRB2005,FratiniPRB2009},
two particle self-consistent approach~\cite{DavoudiPRB2006},
density matrix renormalization group~\cite{VojtaPRB1999,VojtaPRB2001}.
as well as dynamical mean-field
theory~\cite{PietigPRL1999,TongPRB2004,CamjayiNP2008,AmaricciPRB2010,MerinoPRL2013}
(DMFT) and its
extensions~\cite{MerinoPRL2007,Ayral2012PRL,Ayral2013PRB,HuangPRB2014,HafermannPRB2014,LoonPRB2014}.

A seminal study of the EHM within DMFT has been reported in
Ref.~\onlinecite{PietigPRL1999} in the quarter-filling case (i.e. for a total
density $n=0.5$). Using a combination of numerical tools it was
demonstrated the
existence, at large values of the non-local interaction, of a
charge-ordered phase separated from the Fermi liquid metal at weak coupling.
The origin of such symmetry broken state was
interpreted in terms of the effect of strong correlation, signalled by
the enhancement of the effective mass.
For filling smaller than $n=0.5$ and specific values of the local and
non-local interaction the occurrence of phase-separation
between the Fermi liquid and the ordered phase has also been addressed
in Ref.~\onlinecite{TongPRB2004}.
However, the existence of a genuine Mott driven Wigner insulating
state has been demonstrated only later~\cite{CamjayiNP2008} in the
quarter-filling regime. The study of the finite temperature
phase-diagram revealed the existence of a strongly correlated
charge-ordered metal, separating the Wigner-Mott insulator from the
Fermi liquid phase~\cite{CamjayiNP2008,AmaricciPRB2010}.
The existence of a $T=0$ charge-ordered metallic state at quarter-filling has been also shown
using cluster extension of DMFT (CDMFT), which takes into account the
role of short-ranged spatial correlation~\cite{MerinoPRL2007}. In
particular the onset of charge-order was shown to be concomitant with
the occurrence of short range anti-ferromagnetism.
More recently, the effects of non-local interaction has been
studied in two- and three-dimensional cubic lattice using a
combination of GW and Extended DMFT (GW+EDMFT) approach~\cite{Ayral2012PRL,Ayral2013PRB,HuangPRB2014}.
This approach enabled a systematic investigation of the screening
effect and the role of longer-range interaction, up to the third
nearest neighbor~\cite{HuangPRB2014}. The solution the EHM by means of GW+EDMFT
contributed to clarify the phase-diagram of this
model at and near the half-filling regime ($n=1$)~\cite{HuangPRB2014},
associating the interplay between charge-ordering and correlation to
changes in the screening modes.

The emergence of charge-order in strongly correlated systems is also
largely affected by geometric frustration factor, which in turn plays
a relevant role in different systems, e.g. the charge-transfer salts
$\theta$-(BEDT-TTF)$_2$X or the dichalcogenide 1$T$-TaS$_2$ both characterized by a
triangular-lattice geometry.
In this context the interaction driven charge-ordered metal at
quarter-filling is associated to the emergence of a quantum phase,
i.e. pinball liquid. This state is characterized by quasi-localized
charges coexisting  with more itinerant electrons, which gives rise to
strong quasi-particles renormalization with a mechanism analogous to
the heavy fermion compounds~\cite{MerinoPRL2013}. In the multi-orbital
case~\cite{FevrierPRB2015}, which is relevant for transition-metal oxides,  the onset of a
pinball phase has been associated to the a finite value of the Hund's
exchange~\cite{RalkoPRB2015}.

Motivated by the experimental findings and the increasing theoretical
work to understand the nature of strongly correlated electronic
phases in presence of charge ordering, in this paper we investigate
the ground state properties of the EHM. We solve the model using DMFT
with zero temperature Lanczos-based exact diagonalization algorithm.
We present a thorough investigation of the evolution of the
phase-diagram as a function of the local and non-local interaction for
an arbitrary occupation of the system.
We unveil the properties of the
different phase-transitions among the multiple phases characterizing the phase-diagram of
the system, in a full range of variation of the model parameters.
In particular, we address the first- or second-order
nature of the transitions separating the charge-ordered
states from normal (disordered) phases. We study the behavior
of the order parameter and other relevant quantities including the
evolution of the spectral functions.
By evaluating the grand-canonical potential, we also determine the
meta-stability of the solutions across the phase-transitions and
unveil the existence of phase-separation in the phase-diagram.

The rest part of this paper is organized as follows. In Sec.~\ref{sec:modelmetod}, we
introduce the model and the method of solution. We also discuss the
solution of the model in two limiting case.
In Sec.~\ref{sec:phasediagrams} we briefly present the half-filling $n=1$ solution of the
model and discuss the evolution of the phase-diagram as a function of
the chemical potential for an increasing value of the interactions.
In Sec.~\ref{sec:details} we present a detailed analysis for each of the multiple
phase-transitions occurring in the system.
Finally in Sec.~\ref{sec:conclusions} we summarize the results of this work and provide
some future perspectives.


\section{Model and method}\label{sec:modelmetod}

We consider the EHM, which describes the effects of the Coulomb
repulsion onto conduction band electrons in terms of a local and
a nearest-neighbour density-density interaction.
The model Hamiltonian reads:
\begin{equation}\label{eq:model_EHM}
\hat{H}\! =\! -t\!\sum_{\langle
  i,j\rangle,\sigma}\hat{c}^+_{i\sigma}\hat{c}_{j\sigma}\!
+\! U\!\sum_{i}\hat{n}_{i\uparrow}\hat{n}_{i\downarrow}
\!+\! \frac{W}{2}\!\sum_{\langle i,j\rangle} \hat{n}_i\hat{n}_j \!-\! \mu\! \sum_i\hat{n}_i
\end{equation}
where $\langle i,j\rangle$ indicates summation over nearest-neighbor
(NN) sites independently.
The parameter $t$ is the hopping amplitude,
$\hat{c}^{+}_{i\sigma}$ ($\hat{c}_{i\sigma}$) denotes the creation
(destruction) operator of an electron of spin
$\sigma=\uparrow,\downarrow$ at the site $i$.
The operators \mbox{$\hat{n}_{i}=\sum_{\sigma}{\hat{n}_{i\sigma}}$},
\mbox{$\hat{n}_{i\sigma}=\hat{c}^{+}_{i\sigma}\hat{c}_{i\sigma}$},
denote the occupation number. Finally  $\mu$ is the
chemical potential.
The Hamiltonian terms proportional to $U$ and $W$ describe, respectively, the
local and the non-local (NN) part of the screened Coulomb
interaction.  The competition between these two terms enables to
capture the interplay of strong correlation with charge-ordering
effects.

\paragraph*{DMFT solution --} We study the solution of the model (\ref{eq:model_EHM})
using DMFT~\cite{Georges1996RMP}.
In order to allow for a long-range charge ordered phase (check-board
type) the lattice is divided into two sub-lattices, indexed by
$\alpha\!=\!A,B$.
The local nature of the DMFT approach does not allow to accurately
describe non-local interaction terms~\cite{Ayral2012PRL,Ayral2013PRB}. To overcome this problem we
treat the $W$ term at the mean-field level. The corresponding
decoupled Hamiltonian reads:
\begin{eqnarray}\label{eq:model_MFACO}
\hat{H}_{\textrm{MF}} & = &  -t \sum_{\langle i,j\rangle,\sigma}\hat{c}^+_{i\sigma}\hat{c}_{j\sigma} \\
& + & \sum_{\alpha=A,B}\sum_{i\in\alpha}\left[
  U\hat{n}_{i\uparrow}\hat{n}_{i\downarrow} -
\left(\mu-Wn_{\bar{\alpha}}\right)\hat{n}_i \right] + C, \nonumber
\end{eqnarray}
with $C\!=\!-\tfrac{1}{2}W(n^2-\Delta^2)$ an inessential constant
term.

To fix ideas and to further simplify the treatment we shall consider
the case of the Bethe lattice, i.e. a Cayley tree with coordination number
$z\rightarrow+\infty$. The DMFT approximation becomes exact in this
limit, provided $t$ and $W$ are rescaled, respectively, as
$t\rightarrow t/\sqrt{z}$ and  $W\rightarrow W/z$~\cite{Metzner1989PRL}.
In the same limit the decoupling of the
$W$ term is exact, since only Hartree diagram survives \cite{M1989}.
The Bethe Lattice is characterized by a semi-elliptic density of states (per spin):
\begin{equation}\label{eq:DOSBethe}
\rho_0(\epsilon)=\frac{2}{\pi D^2} \sqrt{1 - \left( \epsilon/D \right)^2}\quad \textrm{for}\quad |\epsilon|<D.
\end{equation}
Where $D=2t$ is the  half-bandwidth (or typical kinetic energy)
$D\!=\!2\sqrt{\int^{+\infty}_{-\infty}\epsilon^2\rho_0(\epsilon)d\epsilon}$.
In the following we set $D=1$ as the energy unit.

Within DMFT the quantum many-body lattice problem
(\ref{eq:model_MFACO}) is mapped onto two distinct
effective impurity problems, one per sub-lattice.
The effective baths are described in terms
of frequency dependent Weiss Fields $\bm{\mathcal{G}}^{-1}_0(i\omega_n)\! =\!
\mathrm{Diag}\left[{\cal G}^{-1}_{0A}(i\omega_n),{\cal G}^{-1}_{0B}(i\omega_n)\right]$,
which are self-consistently
determined by requiring to the impurity problems to reproduce the
local physics of the lattice system. In this framework
the self-energy matrix $\bm{\Sigma}(i\omega_n)$, describing the effects of interaction at the
one-particle level, is approximated by its local part. For each
sub-lattice the self-energy function is determined
by the solution of the corresponding quantum impurity problem.
In terms of the matrix self-energy the DMFT self-consistency
condition reads:
\begin{equation}
\bm{\mathcal{G}}^{-1}(i\omega_n) = \bm{G}_\mathrm{loc}^{-1}(i\omega_n)
+ \bm{\Sigma}(i\omega_n)
\label{self-consistency}
\end{equation}
where $\bm{G}_\mathrm{loc}$ is the diagonal matrix of the local interacting lattice Green's
function.
This equation relates the bath properties, expressed by the
Weiss Field, to the local physics of the lattice problem.
In terms of the lattice density of states the components of the local
Green's function are:
\begin{eqnarray}
G_{loc,\alpha}(i\omega_n)  =  \zeta_{\bar{\alpha}}(i\omega_n) \int
\frac{\rho_0(\epsilon) d\epsilon}{ \zeta_A(i\omega_n) \zeta_B(i\omega_n)
  - \epsilon^2},
\label{eqGloc}
\end{eqnarray}
where $\alpha\!=\!A,B$ and $ \zeta_{\alpha}(i\omega_n) = i \omega_n + \mu - W\langle
\hat{n}_{\bar{\alpha}} \rangle - \Sigma_\alpha(i\omega_n)$.

In this work we solve the effective quantum impurity problem using the
Exact Diagonalization technique at $T=0$ \cite{Caffarel1994PRL,Weber2012PRB}.
The effective bath is discretized
into a finite number $N_b$ of levels.
The ground-state of the corresponding Hamiltonian as well as the
impurity Green's functions are determined using the Lanczos technique.
Throughout this work we use $N_b=9$.
The whole DMFT algorithm proceeds as follow. For a given
bath function
$\bm{\mathcal{G}}_0^{-1}$ the effective quantum impurity problems are
solved. The resulting self-energy function $\bm{\Sigma}$ is used to
determine the local interacting Green's function
$\bm{G}_\mathrm{loc}$ by means of Eq.~(\ref{eqGloc}). Finally, using the self-consistency relation
Eq.~(\ref{self-consistency}), a new updated Weiss Field is obtained.
The self-consistent DMFT equations are solved iteratively until the
convergence is reached, usually in few tens iterations. A critical
slowing down of the convergence can be observed near a
phase-transition.

In the following we restrict our attention to the case $U\geq 0$ and
$W\geq0$ and to the {\it non-magnetic} phases.
Furthermore, the model (\ref{eq:model_EHM}) exhibits particle-hole
symmetry (for any $U$ and $W$), so  our results for $n\leq1$
($\bar{\mu}<0$) are holds identically also for $n>1$ ($\bar{\mu}>0$) (see, e.g. Ref.~\onlinecite{MicnasRMP1990}).
Finally, we checked that the occurring phases in this work are solutions with the lowest
grand canonical potential (per site).

\begin{figure}
  \centering
  \includegraphics[width=\rozmiarjeden]{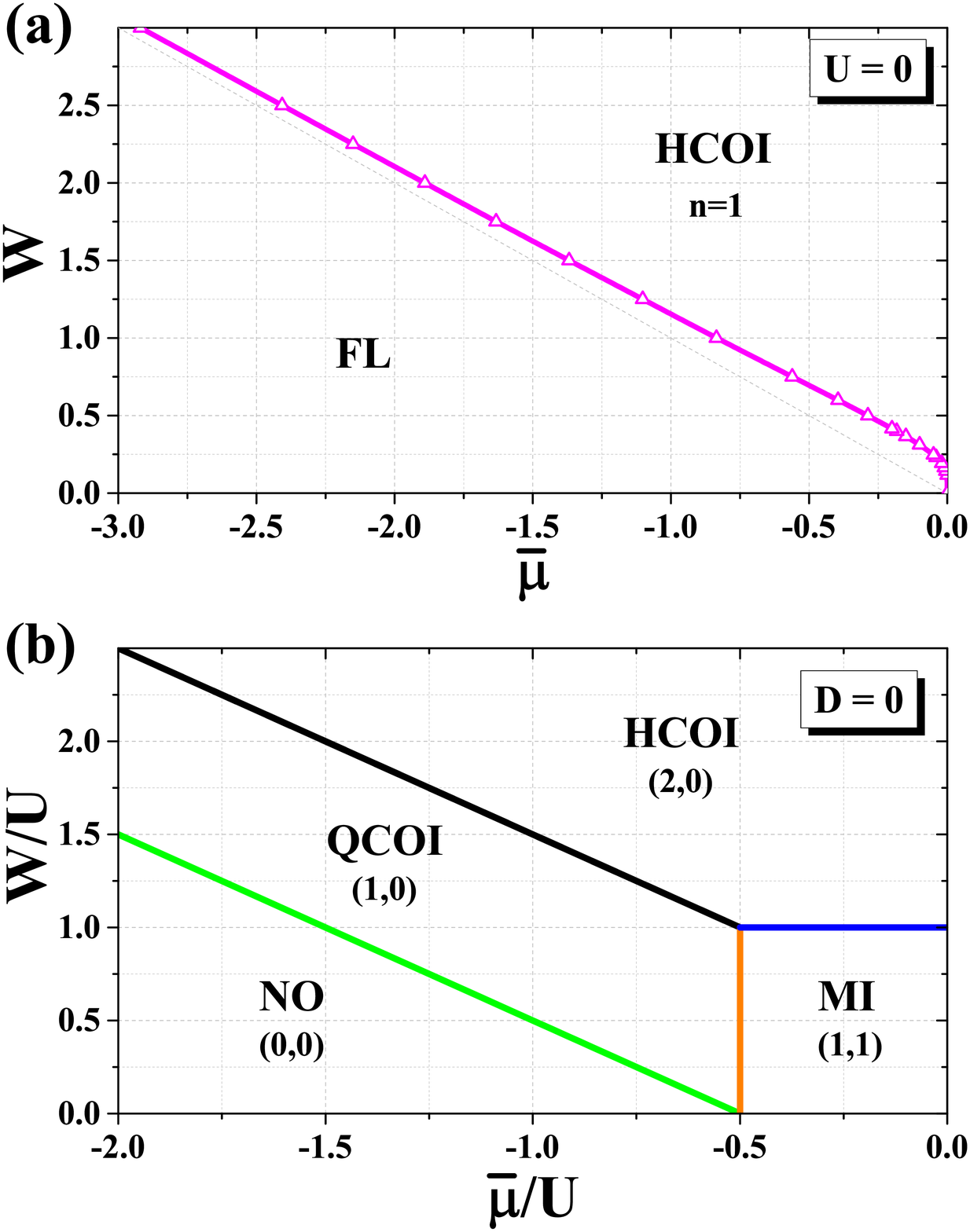}
  \caption{(Color online)
    (a) Phase diagram for $U=0$ as a function of $W$ and
    $\bar{\mu}=\mu-W/2$.
    The boundary between Fermi Liquid (FL) and half-filled charge-ordered insulator ({\HFCOI}) is discontinuous.
    (b) Phase diagram for $t=0$ (atomic limit) as a function of the
    interaction ratio $W/U$ and $\bar{\mu}/U$, $\bar{\mu}=\mu-U/2-W$.
    Each phase is labeled by values of $(n_A,n_B)$.
    All the phases are insulating and all the boundaries are discontinuous.
  }
  \label{rys:GSU0}
\end{figure}


\subsection{Non-interacting limit}\label{sec:noniteractingcase}

For $U=0$ ($W>0$) the model (\ref{eq:model_MFACO}) is solved
using a standard broken-symmetry Hartree-Fock approximation. The
mean-field Hamiltonian is diagonalized by means of a Bogoliubov transformation
\cite{RobaszkiewiczPSSB1973,RobaszkiewiczAPPA1974,VUC1992,UV1993,UV1993PRL,UV1995,CRT2008,CGNR2012a,CGNR2012b}. The resulting self-consistent
equations for the total occupation $n$ and the charge polarization
$\Delta\!=\!\tfrac{1}{2}\left(n_A\! -\! n_B \!\right)$ are:
\begin{eqnarray}
n & = & \frac{1}{2}\int_{-D}^{D} d\epsilon
\rho_0(\epsilon)\left[ f (E_1(\epsilon)) + f (E_2(\epsilon))\right], \\
\frac{\Delta}{W} & = & \Delta \int_{-D}^{D}d\epsilon
\rho_0(\epsilon) \left[ \frac{f(E_1(\epsilon)) - f (E_2(\epsilon))}{2Q(\epsilon)} \right],
\end{eqnarray}
where $f(x)$ is the Fermi function,
$E_{1,2}(\epsilon)\!=\!Wn\!-\!\mu\mp Q(\epsilon)$
and $Q(\epsilon)\!=\!\sqrt{W^2\Delta^2+\epsilon^2}$ \cite{RobaszkiewiczPSSB1973,RobaszkiewiczAPPA1974}.
The grand canonical potential has the form:
\begin{equation}
\Omega  =  \bar{C}
-  \frac{1}{\beta} \sum_{\alpha=1,2} \int d\epsilon \rho_0(\epsilon) \ln\left[ 1\! +\! \exp\left(-\beta E_\alpha(\epsilon)\right) \right],
\end{equation}
with $\bar{C} = -\tfrac{1}{2}W(n^2-\Delta^2)$.

The non-interacting ground-state phase diagram of model (see
Fig.~\ref{rys:GSU0}(a)) shows the existence of two phases, namely
a Fermi liquid metallic state (FL) and a
half-filled, i.e. $n=1$, charge-ordered insulating (\HFCOI).
The transition between these two states is of
first-order, characterized by a discontinuous change of both the occupation $n$
and the polarization $\Delta$ and
phase separation for a definite range of $n$.
For $U\!=\!0$ any $n\!\neq\!1$ charge-ordered metallic (COM) phase is not
stable, i.e. $\partial n / \partial \mu\!<\!0$~\cite{VUC1992,UV1993,UV1993PRL,UV1995,CRT2008,CGNR2012a,CGNR2012b}.
In this limit model (\ref{eq:model_EHM}) is equivalent with the spinless fermion model~\cite{VUC1992,UV1993,UV1993PRL,UV1995,CRT2008,CGNR2012a,CGNR2012b}.


\subsection{Atomic limit}\label{sec:atomiclimit}

In the  $t\!=\!0$ limit the model~(\ref{eq:model_MFACO}) has been
studied in great details (see for example
Refs.~\onlinecite{MRC1984,MM2008,MM2008,PG2008,KR2011,MPS2013,KR2016}
and references therein).
Here, we review briefly the rigorous results in the limit
$z\rightarrow+\infty$ obtained using a variational
approach, which treats the on-site $U$ interaction exactly
and the intersite $W$ interactions within the mean-field approximation
(MFA)~\cite{MRC1984,KR2011,KR2016}.
The ground-state phase diagram for $t=0$ is reported in
Fig.~\ref{rys:GSU0}(b).

The diagram features different COI phases: a  quarter-filling one
(\QFCOI) ($n\!=\!0.5$) and a {\HFCOI} solution.
In addition two non-ordered phases are present: a
band-insulator  for $n\!=\!0$ and a Mott insulating phase at $n\!=\!1$.
Notice that although all transitions are discontinuous, the
homogeneous phases occurring for fixed $n\!\neq\!0.5$ or $n\!\neq\!1$ are degenerated
with phase separated states. Finite temperature or longer-range intersite
interactions however can remove this degeneracy~\cite{KR2011,KR2016}.

\begin{figure}
  \centering
  \includegraphics[width=\rozmiarjeden]{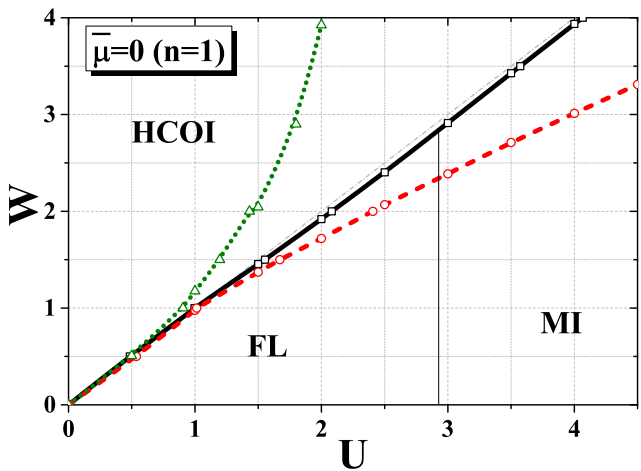}
  \caption{(Color online)
    Phase diagram in the $W$-$U$ plane at half-filling,
    i.e. $n\!=\!1$.
    The solid line (black) at $W/U\!\simeq\!1$ denotes the first-order
    transition separating the half-filled charge-ordered insulator ({\HFCOI}) from the non-ordered phase. The latter
    includes a Fermi Liquid (FL) metal and a Mott Insulator (MI).
    The dashed line (red) delimites the region of meta-stability of the
    {\HFCOI} phase. Similarly the dotted line (green) indicates the region of
    meta-stability for the non-ordered phases.
  }
  \label{fig:GSdiagram.n1}
\end{figure}


\section{Phase diagram}\label{sec:phasediagrams}

We now turn our attention to the combined effect of the local and
non-local interaction in the model (\ref{eq:model_MFACO}).
We first investigated the case of the half-filling,
i.e. $\bar{\mu}=0$ and $n=1$. The phase-diagram in the plane $U$-$W$ is reported in
Fig.~\ref{fig:GSdiagram.n1}. The figure shows the existence in this
regime of three distinct solutions. A {\HFCOI} is found for
$U\!\lesssim\! W$ separated from the normal (non-ordered) solution by a boundary
line just below $W/U=1$. The normal solution includes a FL metal at small
$U$ and a MI for large enough $U$.
In our approach the FL--MI transition line is roughly independent on
$W$. More accurate calculations, taking into account the non-local
interaction beyond the mean-field level, have pointed out that a weak
dependence of the Mott transition line on  $W$~\cite{Ayral2013PRB}.
In the same diagram of Fig.~\ref{fig:GSdiagram.n1} we also denote the
region of meta-stability of the ordered and normal phases enclosed
within two spinodal lines.
While the {\HFCOI} phase extends little into the normal region, we observe
that the coexistence region of the non-ordered solution with the {\HFCOI}
phase rapidly grow with increasing $W$ already for small values of
$U$.

\begin{figure}
	\centering
	\includegraphics[width=\rozmiartrzy]{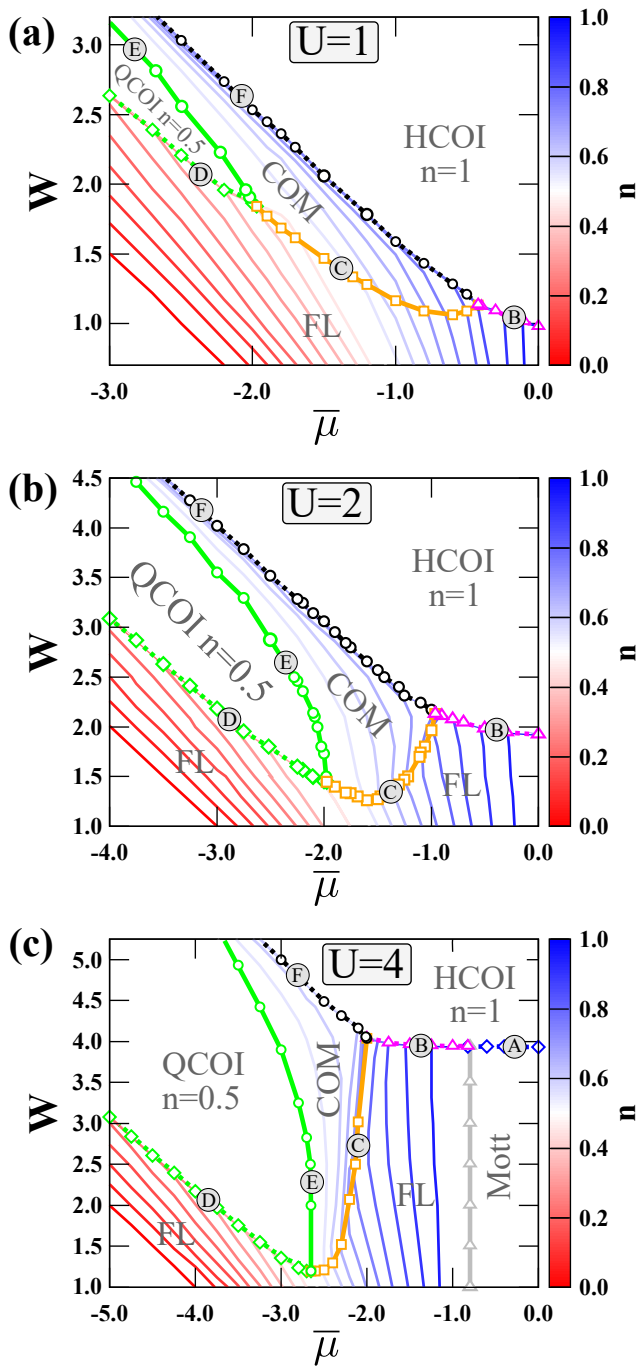}
	\caption{(Color online)
		The $T=0$ phase diagrams in the $W$-$\bar{\mu}$ plane for
		an increasing value
		of the local interaction:
		$U\!=\!1,2,4$ (as labeled).
		The diagrams show the existence of a Fermi Liquid (FL) metallic
		state at small $W$, of a charge-ordered metal (COM) for incommensurate
		occupation as well as different insulating states: a
		quarter-filled ({\QFCOI}), a half-filled charge-ordered
		insulator ({\HFCOI}), and  a Mott insulating phase
		near the $\bar{\mu}\!=\!0$ point.
		The solid lines (C and E) corresponds to a continuous,
		i.e. second-order, transitions while the
		dotted lines (A, B, D and F) indicate the first-order character of the
		phase-transitions. The letters associated to each boundary line
		correspond to the paragraphs in Sec.~\ref{sec:details}.
		The lines within each diagram are iso-density lines, colored according to the
		value of the total density $n$ in the right column.
	}
	\label{fig:GSdiagrams.mi}
\end{figure}

We shall now study the competition of strong correlation and
charge-ordering at finite values of the chemical potential
$\bar{\mu}$.
Our main result is summarized in the phase-diagrams in the plane $W$-$\bar{\mu}$,
reported in Fig.~\ref{fig:GSdiagrams.mi}.
The figure shows the evolution of the diagrams upon increasing the
local correlation strength $U$.
For a finite value of $U$ we observe the presence of two additional phases
with respect to the non-interacting regime (cf. Fig.~\ref{rys:GSU0}(a)),
namely a charge-ordered metal (COM) and a quarter-filled
charge-ordered insulator ({\QFCOI}), i.e. $n\!=\!0.5$ (Figs.~\ref{fig:GSdiagrams.mi}(a) and \ref{fig:GSdiagrams.mi}(b)).
These two phases generically separate the
{\HFCOI} solution from the Fermi liquid metallic state.
A change in the chemical potential
$\bar{\mu}$  first destabilizes the {\HFCOI} towards the charge
ordered metal, occurring at filling $n\!<1$ and for intermediate values of the non-local interaction
$W$. Further increasing the chemical potential the system reaches the
second charge-ordered insulating state, i.e. the {\QFCOI}.

The slope of the boundary line of the {\HFCOI} is governed by the width of the
symmetry related gap, which is linear in $W$.
Upon increasing the strength of the local correlation we observe a
substantial modification of the COM and the {\QFCOI} regions, with the
latter increasing its extension. This is in agreement with the fact
that for larger values of $U$ it becomes easier for the depleted
system to pin the occupation of a single sub-lattice to a commensurate
value, which ultimately favours the formation of a charge-ordered
solution.
Correspondingly the {\HFCOI} region moves towards higher values of
$W$, with the FL-{\HFCOI} transition always occurring at $U\!\simeq\!W$.
Interestingly, the evolution of the boundary line separating the two
metallic phases (Fermi liquid and charge-ordered) is not
monotonous in $U$ (see Fig.~\ref{fig:GSdiagrams.mi}). The negative
slope of the boundary line in the weak interaction regime (Fig.~\ref{fig:GSdiagrams.mi}(a)) is
progressively transformed into a large positive one at strong
coupling (Fig.~\ref{fig:GSdiagrams.mi}(c)).
This change follows directly the evolution of the total
occupation behavior as outlined by the iso-density lines in the
diagrams of Fig.~\ref{fig:GSdiagrams.mi}. By increasing the strength of
the local interaction $U$, the occupation near the FL-COM boundary
line  becomes nearly independent on $W$. This behavior is associated
to the more localized nature of the metallic state (at small doping) near the Mott
insulating phase, occurring for $U\!\gtrsim\!2.92$ near the
$\bar{\mu}\!=\!0$ point (see Fig.~\ref{fig:GSdiagrams.mi}(c)).


\section{Phase transitions}\label{sec:details}

In this section we discuss the properties of
the transitions among the different phases of the system.
To this end, we shall introduce other distinctive quantities, beside
the afore mentioned charge polarization $\Delta$, e.g.
the spectral densities at the Fermi level,
$\rho_{\alpha}(0)\!=\!-\tfrac{1}{\pi}\Im{G}_{\mathrm{loc},\alpha}(\omega\!=\!0)$,
and the renormalization constants  $Z_{\alpha}=\left( 1 -
  {\tfrac{\partial\Sigma_{\alpha}(\omega)}{\partial\omega}}_{|_{\omega\!=\!0}}\right)^{-1}$
($\alpha\!=\!A,B$).


\subsection{The MI-{\HFCOI} transition}

\begin{figure}
        \centering
          \includegraphics[width=\rozmiardwa]{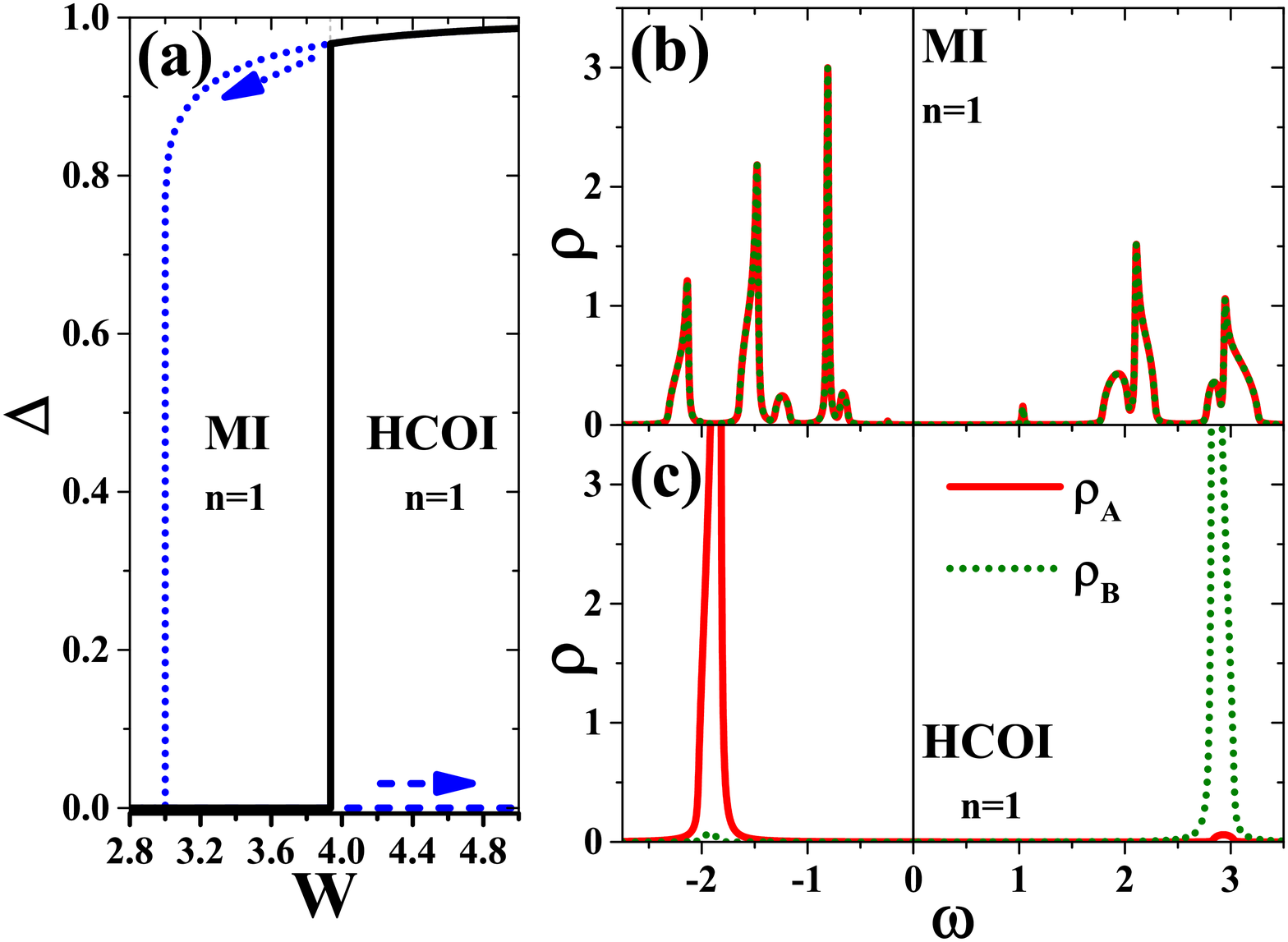}
        \caption{(Color online)
        The behavior of quantities in the neighborhood of the MI-{\HFCOI} boundary.
        (a) $\Delta$ as a function of $W$ for  $U\!=\!4$ and $\bar{\mu}\!=\!-0.5$.
        The solid, dashed, and dotted lines corresponds to stable, metastable MI, and metastable {\HFCOI} solutions, respectively.
        (b) spectral densities for $W\!=\!3.6$ (MI, $n\!=\!1$).
        (c) spectral densities for $W\!=\!4.4$ ({\HFCOI}, $n\!=\!1$).
        Solid and dotted lines corresponds to different sub-lattices (on (b) and (c) panels).
        The Fermi level is at $\omega\!=\!0$.
        }
        \label{fig:GSorderparameters1}
\end{figure}

To start with we discuss the transition between the Mott
insulator and the {\HFCOI} (see Fig.~\ref{fig:GSdiagrams.mi}(c)).
This transition occurs between two phases at half-filling ($n\!=\!1$) and for a large
enough value of $U$ in order to guarantee the existence of the Mott
insulator.
The behavior of the charge polarization $\Delta$ as a function of the
non-local interaction strength $W$ is reported in Fig.~\ref{fig:GSorderparameters1}(a).
The evolution of the order parameter exhibits a discontinuity at the transition
point, indicating its first-order nature as expected for a symmetry
related transition between two insulating states of different origin.
Accordingly, the transition shows a remarkable hysteresis of
$\Delta$, associated with the existence of metastable phases on the
both sides of the transition (see Fig.~\ref{fig:GSorderparameters1}(a)).
The metastable Mott insulating region extends to large values of
$W$ even beyond the limit of the figure.

In panels (b)-(c) of Fig.~\ref{fig:GSorderparameters1} we compare the
spectral densities $\rho_{\alpha}(\omega)\!=\!-\tfrac{1}{\pi}\Im{G}_{\mathrm{loc},\alpha}(\omega)$
($\alpha\!=\!A,B$) in the two phases.
In the Mott region the spectral functions are characterized by the two
contributions at high energy, separated by a gap of the order of $U$
hallmark of the Mott nature of the solution.
The asymmetry of the spectrum is associated with the finite value of
the non-local term $W$.

In the {\HFCOI} region (see Fig.~\ref{fig:GSorderparameters1}(c)) the spectral density is
characterized by a band gap, separating the completely filled
sub-lattice $A$ ($n_A\!\thickapprox\!2$) from that of the empty sub-lattice
$B$ ($n_B\!\thickapprox\!0$).
The gap is related to charge-order phenomenon and has a width
of $\simeq\!2(W-U/2)\Delta$. Because of its mean-field nature with
respect to symmetry breaking transition, the DMFT description of
charge-ordered state closely reminds the mean-field solution of the
problem \cite{VUC1992}.

\subsection{The FL-{\HFCOI} transition}

\begin{figure}
        \centering
          \includegraphics[width=\rozmiardwa]{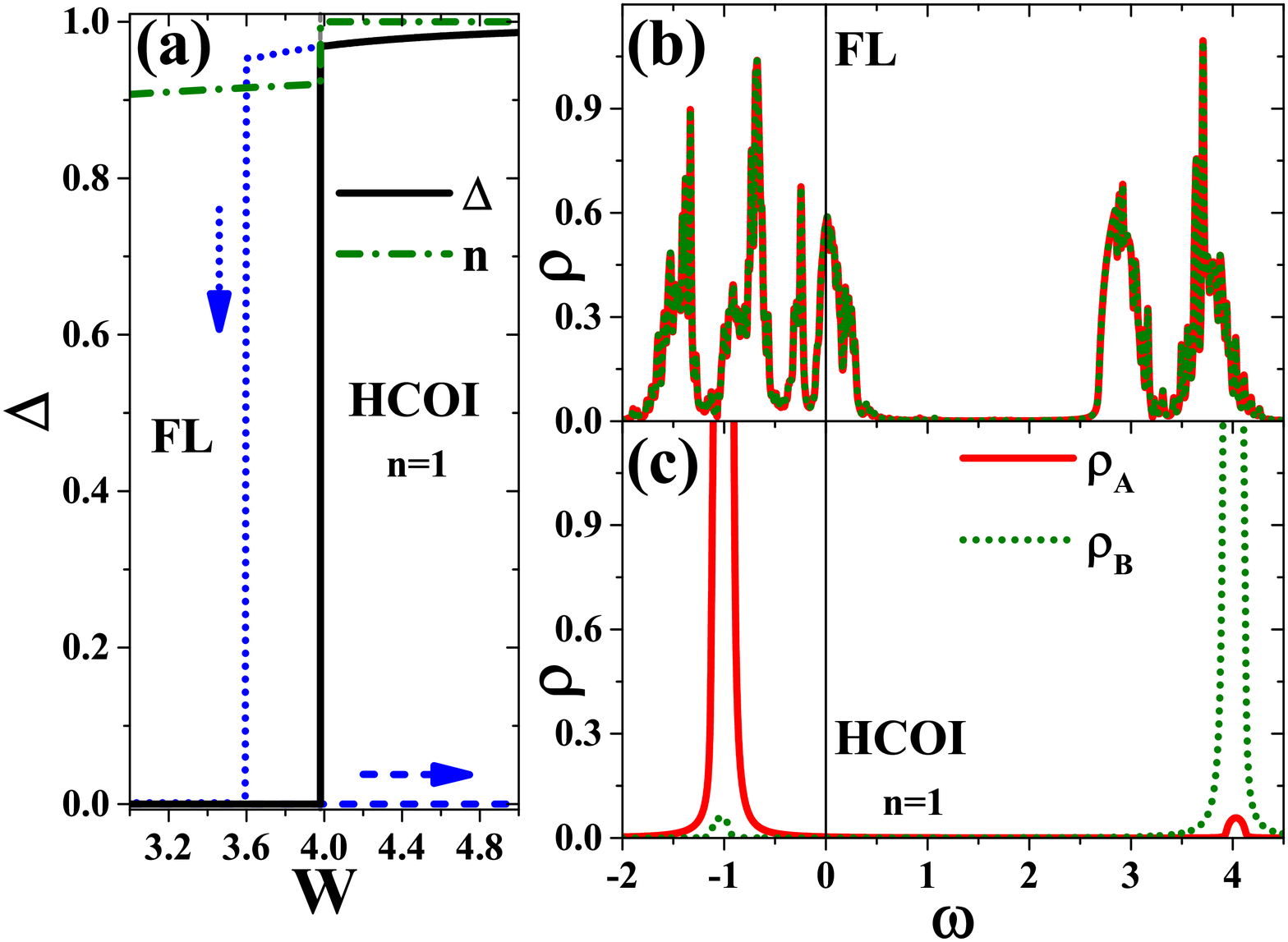}
        \caption{(Color online)
        The behavior of quantities in the neighborhood of the FL-{\HFCOI} boundary.
        (a) $\Delta$ (solid line) and $n$ (dashed-dotted line) as a function of $W$ for $U\!=\!4$ and $\bar{\mu}\!=\!-1.5$.
        The  dashed and dotted lines corresponds to metastable FL and metastable {\HFCOI} solutions, respectively.
        (b) spectral densities for $W/D\!=\!3.6$ (FL).
        (c) spectral densities for $W/D\!=\!4.4$ ({\HFCOI}, $n\!=\!1$).
        Solid and dotted lines corresponds to different sub-lattices (on (b) and (c) panels).
        The Fermi level is at $\omega\!=\!0$.
        }
        \label{fig:GSorderparameters2}
\end{figure}

At finite doping the system admits a transition between the Fermi liquid
metal and the {\HFCOI} solution. Differently from the previous case
this transition occurs between states at different fillings.
The behavior of the order parameter $\Delta$ and the occupation $n$
across the transition boundary are  reported in
Fig.~\ref{fig:GSorderparameters2}(a).
Both quantities exhibit an abrupt change at the transition.
In particular, $n$ evolves discontinuously from $n=1$, in the ordered
phase, to $n\!\thickapprox\!0.92\!<\!1$ in the normal metallic region.
The first-order character of the FL-{\HFCOI} transition is further
underlined by the hysteresis of $\Delta$.
Analogously to the Mott phase, the region of metastability of the FL
phase extends to large values of $W$ (beyond the range of the figure).

The spectral functions of the two solutions for the $U=4$ case are reported in
Fig.~\ref{fig:GSorderparameters2}(b)-(c).
In the FL phase the spectral densities, identical for both sub-lattices
$\rho_A(\omega)\!=\!\rho_B(\omega)$, have a finite weight at the Fermi
level ($\omega\!=\!0$) obtained by small doping ($\bar{\mu}\!=\!-1.5$)
the Mott insulating solution. The  quasi-particle resonance at the Fermi level
flanks the lower Hubbard band which in turn is separated by a gap of order
$U$ from the upper one.
In the panel (c) we show the spectral densities for the {\HFCOI}
phase. The rigid shift with respect to
Fig.~\ref{fig:GSorderparameters1}(c) is due the different value of the
chemical potential. However the properties of this phase remains
unchanged.

\subsection{The FL-COM transition}

\begin{figure}
        \centering
          \includegraphics[width=\rozmiardwa]{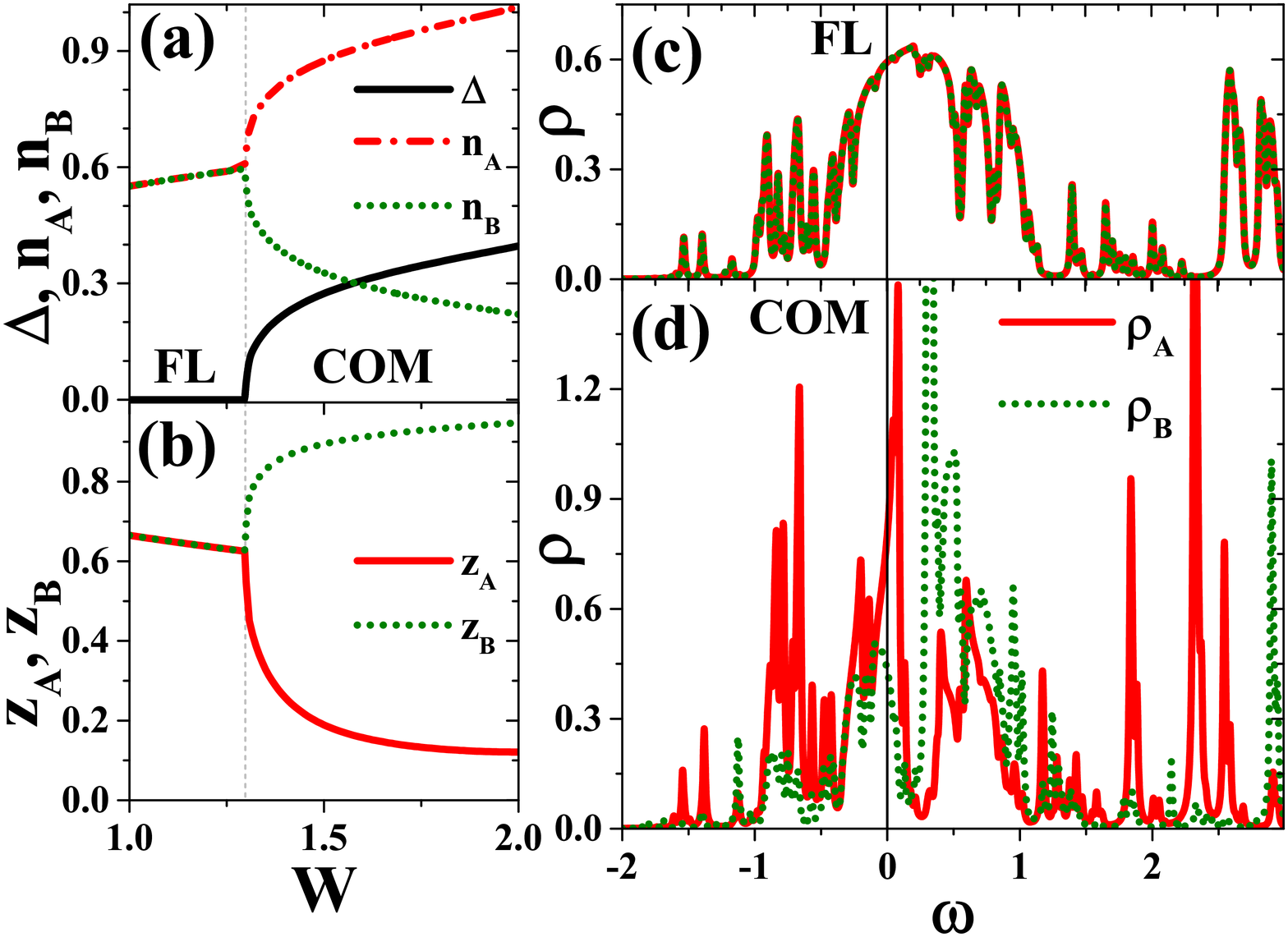}
        \caption{(Color online)
        The behavior of quantities in the neighborhood of the FL-COM boundary.
        (a) $\Delta$ (solid line), $n_A$ (dashed-dotted line), and $n_B$ (dotted line) as a function of $W$ for $U/D\!=\!2.0$ and $\bar{\mu}/D\!=\!-1.5$.
        (b) $z_A$ (solid line) and $z_B$ (dotted line) as a function of $W$ for the same values of the other parameters.
        (c) spectral densities for $W\!=\!1.25$ (FL) .
        (d) spectral densities for $W\!=\!1.35$ (COM).
        Solid and dotted lines corresponds to different sublattices (on (c) and (d) panels).
        The Fermi level is at $\omega\!=\!0$.
        }
        \label{fig:GSorderparameters3}
\end{figure}

Similarly, the FL metal can be destabilized towards a COM phase by either
increasing the non-local interaction $W$ or the doping value. The
symmetry breaking transition relating these two phase at
incommensurate filling is continuous, i.e. of second order.
Further insight in the continuous character of the transition can be
inferred from the behavior of the iso-density lines across the
boundary line, see Fig.~\ref{fig:GSdiagrams.mi}.
Approaching the transition from both phases, the density evolves
smoothly enabling for the continuous formation/destruction of the
charge-polarization.
The behavior of the densities $n_{A,B}$ and of the order parameter
$\Delta$ as a function of $W$ for $U=2$ is reported in
Fig.~\ref{fig:GSorderparameters3}(a).
Crossing the transition line (see the phase-diagram in Fig.~\ref{fig:GSdiagrams.mi})
by decreasing $W$ from the COM solution, the order parameter $\Delta$
gets continuously reduced to zero. In proximity of the critical point
$W\!=\!W_c$ the order parameter exhibits the characteristic
square-root behavior $\Delta\!=\!(W-W_c)^{1/2}$ (see
Fig.~\ref{fig:GSorderparameters3}(a)) as expected from a mean-field
description of the phase-transition.

In the charge-ordered phase the unbalanced occupations among the two
sub-lattices give rise to different degrees of correlation. We
quantify this by showing the behavior of  the renormalization
constants $Z_{A,B}$ across the phase-transition in
Fig.~\ref{fig:GSorderparameters3}(b).
In particular, the  sub-lattice $A$ becomes nearly half-filled
($n_A\!\simeq\!1$)  while the other $B$ gets slightly
depleted. Correspondingly $Z_A\!<\!Z_B$, i.e. the metallic state at
$A$ becomes significantly more correlated than the other, less
occupied, sub-lattice.

The different nature of the metallic states at the two phases is
also evident from the corresponding spectral functions,  reported in
Fig.~\ref{fig:GSorderparameters3}(c)-(d).
A large featureless spectral weight, reminiscent of the
non-interacting (semi-elliptic) distribution, characterizes both
sub-lattices in the FL phase. The effect of the large correlation $U$
manifests itself in the contribution at high-energy (Fig.~\ref{fig:GSorderparameters3}(c)).
In the charge-order phase the two spectral functions show
the formation of a tiny gap slightly away from the Fermi
level ($\omega\!=\!0$), as result of the
incommensurate filling and to the metallic character
of the solution.
In this regime the two sub-lattices distributions are nearly specular
one with respect to another, with a relative shift of about $2W\Delta$.
The strongly correlated nature of the sub-lattice $A$ solution is further
underlined by the narrow resonance at the Fermi level characterizing
the  spectral function at low energy (see
Fig.~\ref{fig:GSorderparameters3}(d)). Likewise, precursors of the
Hubbard bands are visible in the higher energy region.

\subsection{The FL-{\QFCOI} transition}

\begin{figure}
	\centering
	\includegraphics[width=\rozmiardwa]{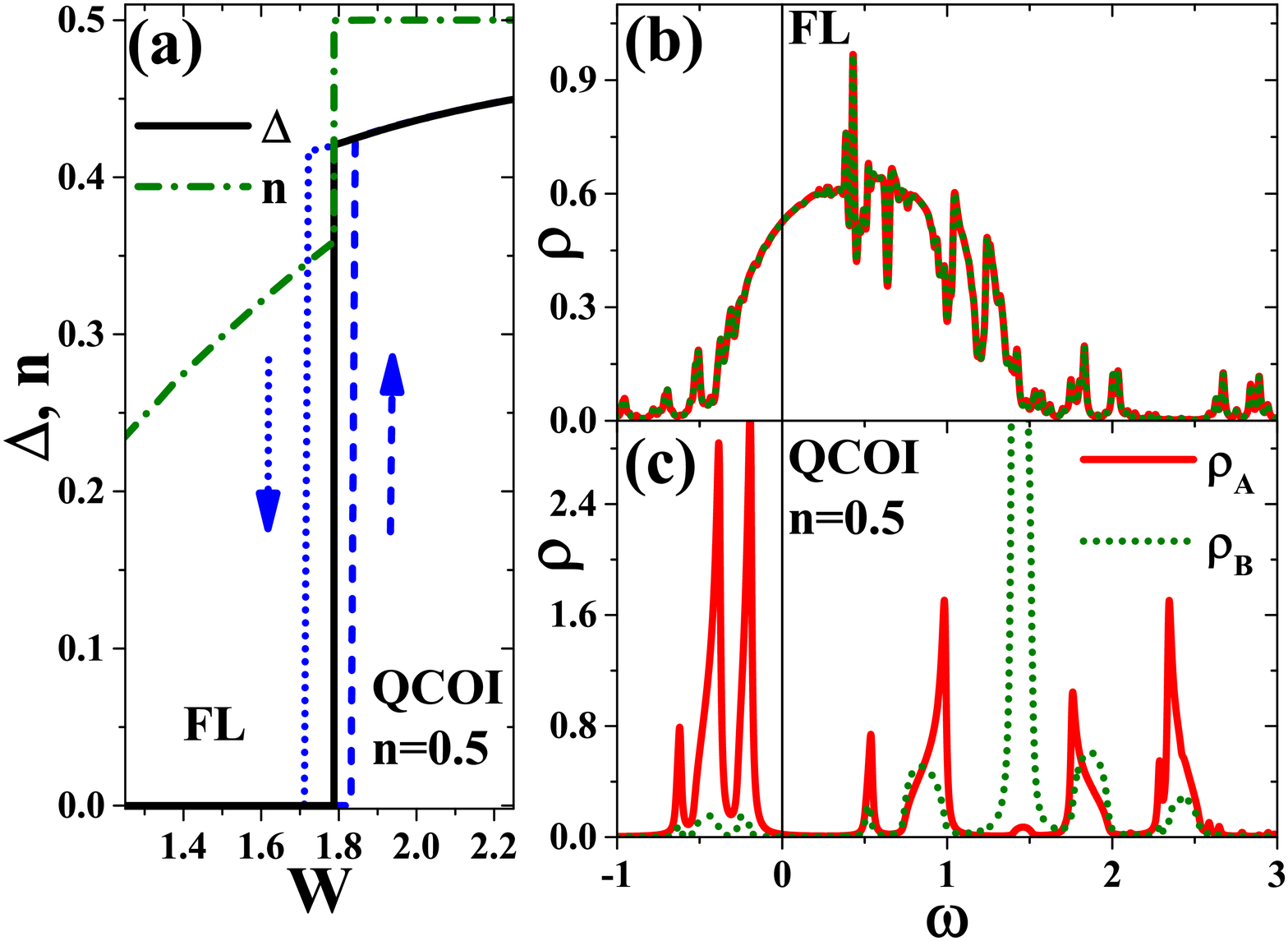}
	\caption{(Color online)
		The behavior of quantities in the neighborhood of the FL-{\QFCOI} boundary.
		(a) $\Delta$ (solid line) and $n$ (dashed-dotted line) as a function of $W$ for $U\!=\!2.0$ and $\bar{\mu}\!=\!-2.5$.
		The  dashed and dotted lines corresponds to metastable FL and metastable {\QFCOI} solutions, respectively.
		(b) spectral densities for $W\!=\!1.70$ (FL).
		(c) spectral densities for $W\!=\!1.85$ ({\QFCOI}, $n\!=\!0.5$).
		Solid and dotted lines corresponds to different sub-lattices (on (b) and (c) panels).
		The Fermi level is at $\omega\!=\!0$.
	}
	\label{fig:GSorderparameters4}
\end{figure}

For even larger values of the chemical potential
$\bar{\mu}$, i.e. doping, the effect of non-local interaction $W$ is to transform the
Fermi liquid metal directly into a {\QFCOI}. The transition occurs
through the pinning of the occupation to a commensurate value for one
sub-lattice with the concomitant opening of a charge-order gap, while the other
sub-lattice gets nearly empty.
The discontinuous nature of the transition can be further appreciated
by looking at the evolution of the iso-density lines near the boundary
line. As reported in Fig.~\ref{fig:GSdiagrams.mi}, near the transition
the FL metal has a small occupation while the {\QFCOI} is pinned to
$n\!=\! 0.5$. This prevents the continuous transformation of one state
into the other and the only way to connect the two phases is through a
first-order jump.

The behavior of the density $n$ and of the order parameter $\Delta$ as
a function of $W$ across the transition is reported in
Fig.~\ref{fig:GSorderparameters4}(a).
The occupation $n$ undergoes a jump to $n\!=\!0.5$
for a critical value of the non-local interaction $W\!=\!W_c$. At the
critical point the charge-polarization $\Delta$ suddenly acquires a finite
value. The figure also shows the hysteresis cycle of the
order-parameter, demonstrating the first-order nature of this
transition.

In the panels (b)-(c) of the figure we report the evolution of the
spectral functions across the phase-transition. In the FL phase (see
Fig.~\ref{fig:GSorderparameters4}(b)), the distribution is
characterized by a large featureless spectrum. In this low-density
regime and for small $W$ the effects of the local correlation are
rather weak.  Entering the {\QFCOI} region both spectral functions
exhibit a gap of the order of $\textrm{Max}\{2W\Delta,U\}$ at the Fermi level,
associated with the charge-order and Mott localization occurring in the system.
The spectral weight of the sub-lattice B is located above the Fermi
level, corresponding to an almost depleted regime, while the other
sub-lattice is nearly half-filled due to intersite repulsion $W$.

The interplay between both $U$ and $W$ interaction has a strong impact
on this phase and, as discussed in Sec.~\ref{sec:phasediagrams} and
Ref.~\onlinecite{AmaricciPRB2010}, they both contribute to stabilize the {\QFCOI} phase.
Thus, the occurrence of the {\QFCOI} phase is associated with the conventional Mott
scenario for the localization of the electrons in nearly half-filled
sub-lattice $A$, whereas almost empty sub-lattice $B$ is rather band
insulator \cite{AmaricciPRB2010}.


\subsection{The {\QFCOI}-COM transition}

\begin{figure}
	\centering
	\includegraphics[width=\rozmiardwa]{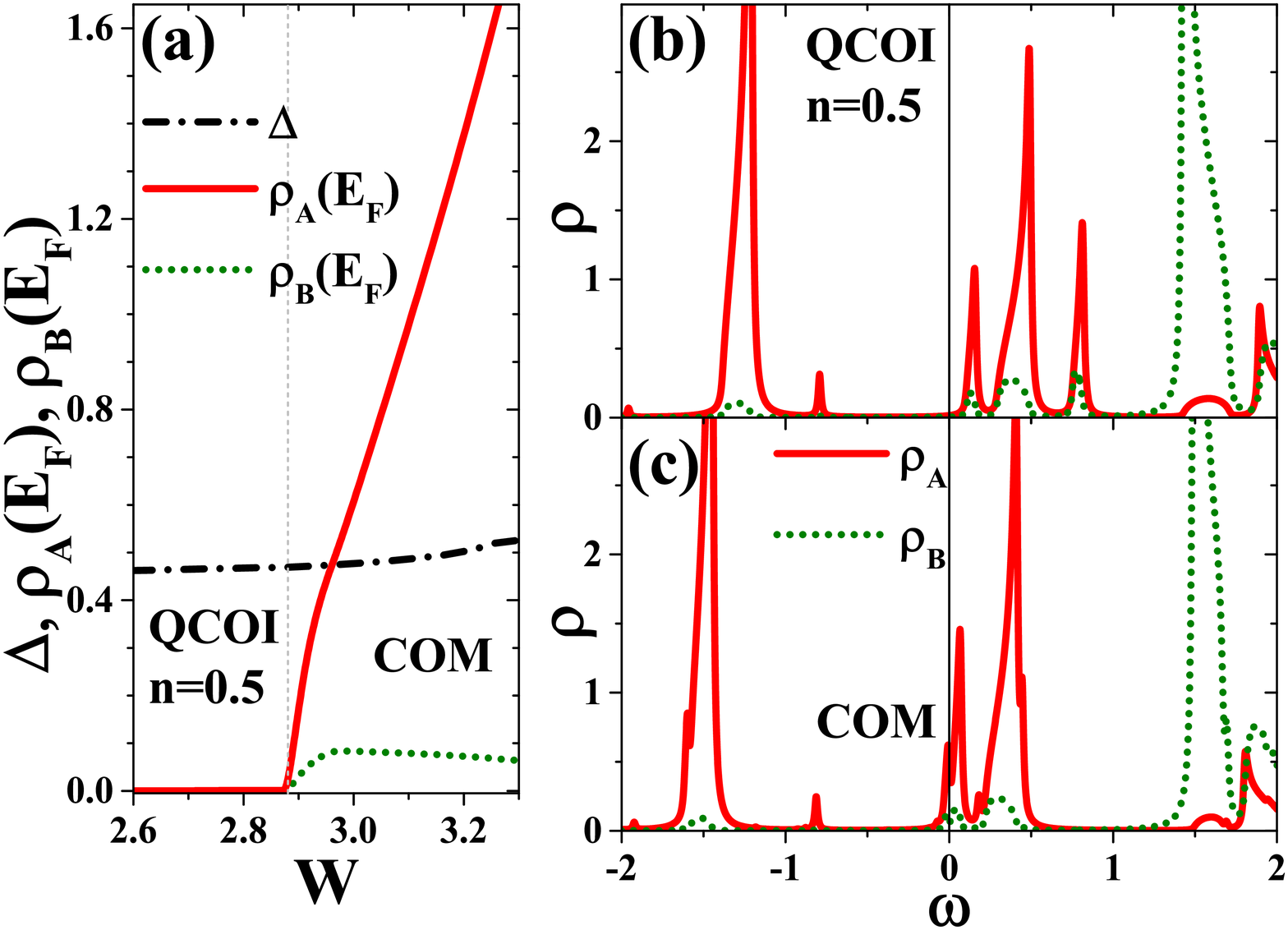}
	\caption{(Color online)
		The behavior of quantities in the neighborhood of the {\QFCOI}--COM boundary.
		(a) $\Delta$ (dashed-dotted line), $\rho_A(E_F)$ (solid line), and
		$\rho_B(E_F)$ (dotted line) as a function of $W$ for $U\!=\!2.0$ and $\bar{\mu}\!=\!-2.5$.
		(b) spectral densities for $W\!=\!2.75$ ({\QFCOI}, $n\!=\!0.5$).
		(c) spectral densities for $W\!=\!3.00$ (COM).
		Solid and dotted lines corresponds to different sublattices (on (b) and (c) panels).
		The Fermi level is at $\omega\!=\!0$.
	}
	\label{fig:GSorderparameters5}
\end{figure}

The charge-ordered insulating phase at quarter filling can be
destabilized by either reducing the doping (i.e. reduce the
chemical potential) or increasing $W$. The resulting insulator to
metal transition however occurs without destroying the long range
charge order.
In Fig.~\ref{fig:GSorderparameters5}(a) we demonstrate this by tuning
the non-local interaction $W$, driving the {\QFCOI} into a COM state.
In this figure we show that charge polarization $\Delta$ remains finite across
the transition.
We characterize the metallization process through the behavior of
the spectral weights at the Fermi level $\rho_{A,B}(E_F\!\equiv\!0)$ across
the transition. These quantities show a continuous evolution,
corresponding to a second-order phase-transition. In particular,
the sub-lattice $A$ which is near the half-filling occupation, rapidly gain a
substantial amount of spectral weight at Fermi level developing
continuously into a strongly correlated metal, i.e. $Z_A\!\simeq\!0$
(not shown in the figure) in agreement with the Wigner-Mott
transition scenario for the electrons at the sub-lattice $A$~\cite{CamjayiNP2008,AmaricciPRB2010}.

The spectral distributions across the transition are reported in
Fig.~\ref{fig:GSorderparameters5}(b)-(c). The two cases are
characterized by the presence of a large charge-order gap. Within our
solution, increasing of the interaction term $W$ cause a small change
of the chemical potential, which is sufficient to destabilize
the charge-order insulator into a metallic state. However, in a presence
of sufficiently strong local and non-local correlation a small doping is not enough
to suppress completely the long range ordering, leading to the
formation of the COM.


\subsection{The COM-{\HFCOI} transition}

\begin{figure}
	\centering
	\includegraphics[width=\rozmiardwa]{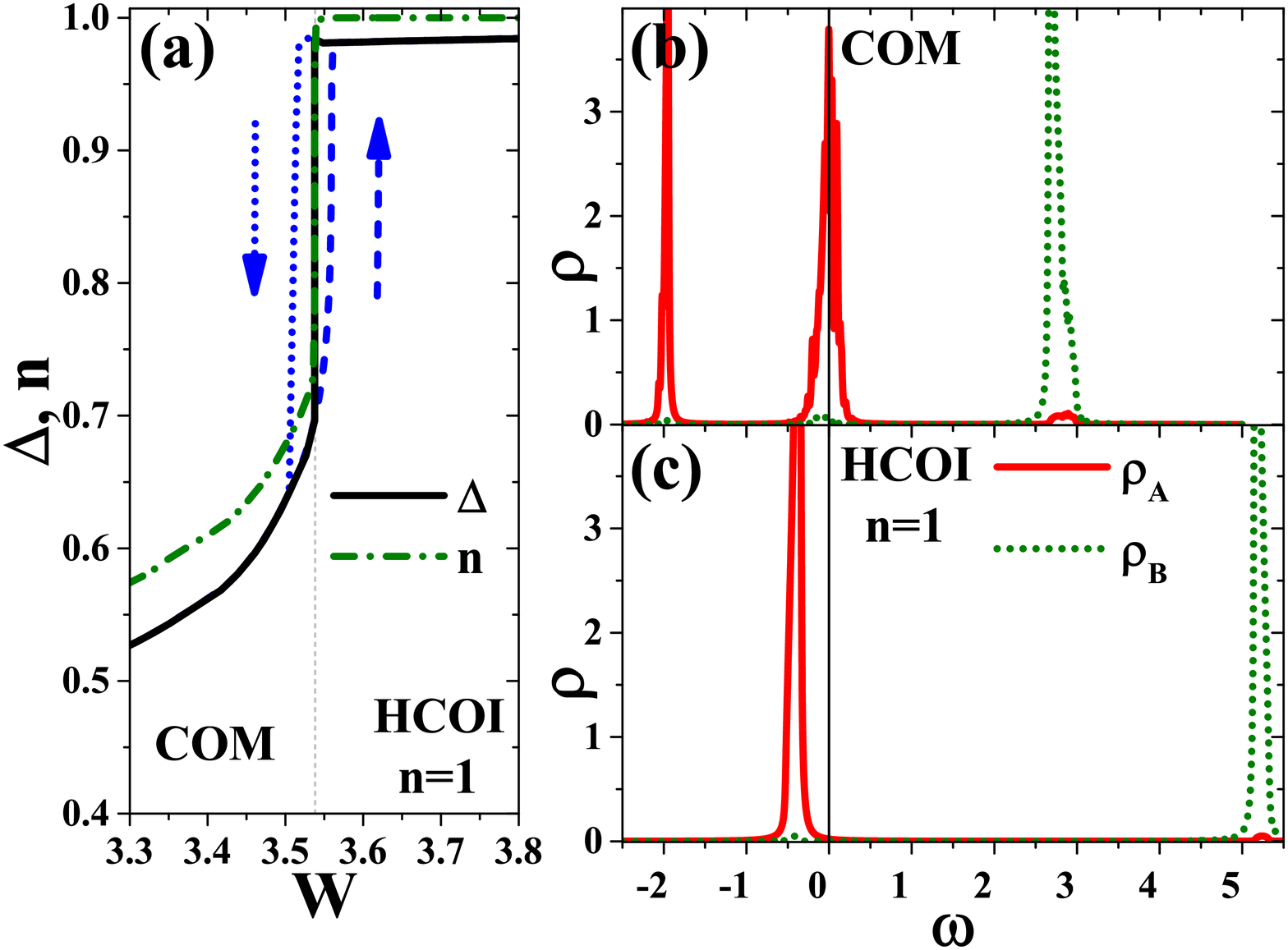}
	\caption{(Color online)
		The behavior of quantities in the neighborhood of the COM--{\HFCOI} boundary.
		(a) $\Delta$ (solid line) and $n$ (dashed-dotted line) as a function of $W$ for $U\!=\!2.0$, $\bar{\mu}\!=\!-2.5$.
		The  dashed and dotted lines corresponds to metastable COM and metastable {\HFCOI} solutions, respectively.
		(b) spectral densities for $W\!=\!3.5$ (COM).
		(c) spectral densities for $W\!=\!3.8$ ({\HFCOI}, $n\!=\!1$).
		Solid and dotted lines corresponds to different sub-lattices (on (b) and (c) panels).
		$\omega=0$ corresponds to the Fermi level.
	}
	\label{fig:GSorderparameters6}
\end{figure}

We finally discuss the properties of the transition from the COM state
to the {\HFCOI} phase.
In Fig.~\ref{fig:GSorderparameters6}(a) we report the behavior as a
function of the interaction $W$ of both the total density $n$ and the
charge-polarization $\Delta$ across the transition line.
Differently from the COM-{\QFCOI} case discussed previously, the
transition from the charge-ordered metallic state to the {\HFCOI} has a
first-order character. As reported in the figure at the critical point
the total density jumps to the $n\!=\!1$ value. Correspondingly, the
order parameter discontinuously reaches its maximum value. The
hysteresis of this quantity is also reported in the figure.

The spectral densities across the transition are shown in Fig.~\ref{fig:GSorderparameters6}(b)-(c).
The evolution of the spectra reveals that the metal-insulator
transition is driven by a shift of the spectral weight below the Fermi
level, i.e. the interaction $W$ drives a shift in the chemical
potential increasing the total occupation and a concomitant transfer of the
spectral weight. In this regime the large
value of $W$ maintains the long range charge order.
The {\HFCOI} spectrum displays the distinctive features already
discussed in the previous sections.


\section{Conclusions}\label{sec:conclusions}

In this work we studied the competition between local and non-local
electronic interaction within the paradigmatic extended Hubbard
model. We solved the model non-perturbatively using dynamical
mean-field theory, with a Lanczos Exact Diagonalization
technique.
In particular we thoroughly investigated the interplay of
charge-ordering and Mott physics as a function of the chemical potential which in turn controls
the particle density.
We determined the zero temperature phase-diagram as a function of the non-local
interaction $W$ and chemical potential $\bar{\mu}$.
For any value of local correlation $U$ we reported the existence of
both an insulating charge ordered solution at quarter filling and a
incommensurate charge ordered metal. These two phases, which have no
counterpart in the non-interacting regime, get stabilized by the
interplay of local and longer range interaction. The evolution of the
phase-diagram as a function of $U$ has shown the increasing stability
of the quarter-filled charge-ordered solution.

We studied in details the nature and the properties of the different
phase-transitions occurring among the multiple phases of the
system. In particular we unveiled the characteristics of the
continuous metal-insulator transition separating the charge-ordered metal and the $n\!=\!0.5$
insulator, which extends to the incommensurate case previous results
available in the recent literature.
Moreover, we showed that the small $W$ Fermi liquid metal is
separated from the charge-ordered metallic state by a continuous
transition and from the quarter-filled charge ordered insulator by a
first-order one.
The analysis of the iso-density lines enabled us to associate the
difference in the transition character to distinct behavior of the
occupation in the two regimes.
Thus, for example the incommensurate filling of the charge ordered metal
can be continuously connected with the Fermi liquid state through the
progressive reduction of the charge polarization. On the contrary
in the large doping regime, the severe difference in the occupation
between the quarter filled charge-ordered insulator and the almost
empty Fermi liquid leaves room only to a first order phase-transition.

Although the simple nature the Extended Hubbard Model cannot be
regarded as a realistic or quantitatively accurate representation of
the real system, its solution in a controlled non-perturbative
framework allows to shed light on the microscopic mechanism
behind several experimental observations.
Notice also that the immense development of experimental techniques
in cold atomic Fermi gases on the optical lattices in the last years has opened
new opportunities  for research of strongly correlated systems and beyond.
The ability to precisely control the interactions via Feshbach resonances \cite{fedichevprl96,partridge.Science.06,zwierlein.Science.06}
sets new perspectives for experimental realization and
study of many different theoretically well-described system,
in particular various non-standard Hubbard models (for review see, e.g., Refs. \onlinecite{georgescu.RMP.14,dutta.rpp.15}).

Our study demonstrates how the tendency to charge ordering favors and strengthens
the transformation of conduction electrons into localized particles in
the presence of long-range charge-order.
The analysis of the phase-transitions and the destruction of the
charge-order at an arbitrary filling can be useful to understand
recent experiments on the 2D electron gas from deposed liquid He$^3$ on
a substrate, which corresponds to incommensurate density~\cite{Andrei1997,HaquePRB2003,KonstantinovPRL2011,LevitinPRL2013,Yayama2014,AoyamaPRB2014}.


\begin{acknowledgments}
K.J.K thanks SISSA for the hospitality during his six month research
stay in Trieste in 2015.
A.A. and M.C. acknowledge financial support from the European Research
Council under FPO7 Starting Independent Research Grant n.240524
``SUPERBAD".
A.A. also acknowledges support from the European Union,
Seventh Framework Programme FP7, under Grant No. 280555 ``GO FAST''
and under H2020 Framework Programme, ERC Advanced Grant No. 692670
``FIRSTORM'' for part of this work.
K.J.K was supported by National Science Centre
(NCN, Poland) -- ETIUDA 1 programme, grant No. DEC-2013/08/T/ST3/00012
in years 2013--2015.
\end{acknowledgments}


\bibliography{bibliography}

\end{document}